\documentstyle[prl,aps,twocolumn,psfig,floats]{revtex}
\input{epsf}
\begin{document}
\preprint
\draft
\twocolumn[\hsize\textwidth\columnwidth\hsize\csname@twocolumnfalse\endcsname
\title{Time Reversal Symmetry Breaking and Spontaneous Currents in\\
        s-Wave / Normal Metal / d-Wave Superconductor Sandwiches}
\author{Andrea Huck$^{a}$, Anne van Otterlo$^{b,*,\dagger}$,
        and Manfred Sigrist$^{c}$}
\address{a) I. Physikalisches Institut, Universit\"{a}t Hamburg,
        D-20355 Hamburg, FRG\\
        b) Physics Department, University of California,
        Davis, CA 95616, USA \\
        c) Theoretische Physik, ETH-H\"{o}nggerberg, CH-8093 Z\"{u}rich,
        Switzerland}
\maketitle

\begin{abstract}
We study the physical properties of an $s$-wave -- normal metal --
$d$-wave junction in terms of the Andreev bound states in the normal
metal layer.  The phase dependence of bound states with different
orientations leads to superconducting states with broken time reversal
symmetry for generic orientations of the $d$-wave superconductor
crystal.  The occurrence of such a state and the associated
spontaneous supercurrent along the junction is analyzed in the
framework of Ginzburg-Landau theory and by the solution of the
Bogoliubov-de Gennes equations.
\end{abstract}

]

During the last few years the order parameter (OP) symmetry has been
one of the intensively debated issues in the field of high-temperature
superconductivity.  A growing number of experiments leaves
little doubt that the basic symmetry of the Cooper pairs has
$d_{x^2-y^2}$-wave character in many of the high-temperature
superconductors (HTSC)~\cite{SCALAPINO,HARLINGEN}.  The unconventional
symmetry of the OP has important implications for the Josephson
effect.  For d-wave superconductors the Josephson coupling is subject
to an additional phase dependence caused by the internal phase
structure of the Cooper pair wave function.  The phase properties of
the Josephson effect have been discussed within the framework of the
generalized Ginzburg-Landau (GL)~\cite{SRREV} as well as the tunneling
Hamiltonian approach~\cite{BOZ}.  It was found that the current-phase
relation depends on the mutual orientation of the two coupled
superconductors and their interface.  This property is the basis of
all the phase sensitive experiments probing the OP symmetry. In
particular, it is possible to create multiply connected d-wave
superconductors which generate half-integer flux quanta as observed in
experiments~\cite{TSUEI}.

Various interesting phenomena occur in $45^o$-interfaces of
$d_{x^2-y^2}$-wave superconductors, where one of the nodes of the pair
wave function lies parallel to the interface normal vector (Fig.~1).
For an interface to a normal metal or an insulator a bound state
appears at zero energy giving rise to a zero-bias anomaly in the
$I$-$V$-characteristics of quasiparticle tunneling~\cite{HU,TANAKA}.
It was also shown that in such an interface to an s-wave
superconductor the energy minimum corresponds to a Josephson
phase different from $ 0 $ or $ \pi $ \cite{YIP}. Based on Ginzburg-Landau
theories it was suggested that this is connected with a local
breakdown of time reversal symmetry ${\cal T}$~\cite{SBL,SK}.  
The s-wave and 
d-wave OP can form a complex combination, a so-called ($s+id$)-state,
close to this $45^o$-junction.  This leads to a phase difference of
$+\pi/2$ or $-\pi/2$ across the interface, which corresponds to two
degenerate states~\cite{SK,MATSUMOTO}.  It can be seen from the GL
formulation that under this condition a spontaneous current flows
parallel to the interface which produces a local field
distribution~\cite{SBL}.

\begin{figure}[hbt]
        \label{fig:geom}
        \unitlength1cm
        \begin{picture}(8.0,7.5)
        \put(0.4,0){\psfig{figure=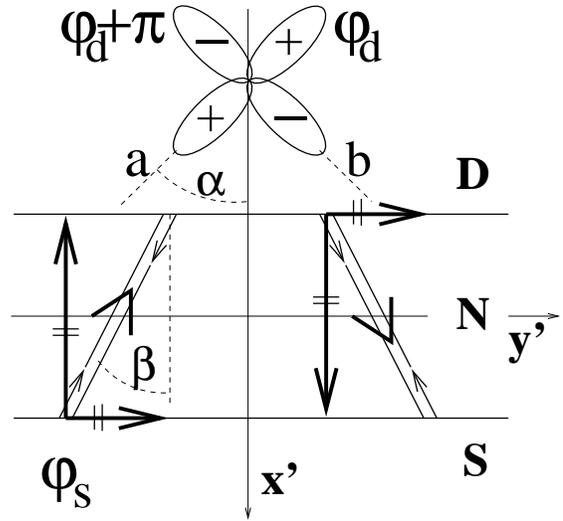,height=7.5cm,width=8.0cm}}
        \end{picture}
\caption{Schematic view of the SND-junction.  The angle $\alpha$
  denotes the orientation of the d-wave superconductor (crystal axis
  $a$ and $b$) and $\beta$ the momentum direction of the bound state.
  The currents generated by the bound states tend to cancel in the
  direction perpendicular to the interface, whereas they add parallel
  to the interface to a spontaneous current.}
\end{figure}

In this paper we consider a $45^o$-interface with a normal metal
between the d-wave and the s-wave superconductor, a device which we
call the SND-junction~\cite{CURACAO}. Also for this configuration a $
{\cal T} $-violating state appears and generates a supercurrent mainly
in the region of the normal metal.  It is our goal to demonstrate that
this current has a simple and intuitive interpretation in terms of
subgap Andreev bound states in the sandwiched normal metal layer. Let
us first outline the basic idea for the situation shown in
Fig.~1 where $\alpha=\pi/4$ and the c-axis is parallel to the
interface.  In terms of the phase difference
$\varphi=\varphi_{d}-\varphi_{s}$, the Josephson current carried by a
bound state with a specific orientation $\beta$ can be expanded in
harmonics of the phase difference $\varphi$ as
$I_{\beta}(\varphi)=I_{1}(\beta)\sin(\varphi)
+I_{2}(\beta)\sin(2\varphi)+\cdots$.  In the geometry considered, each
bound state with orientation $0<\beta<\pi/4$ that sees the ``+'' lobe
with phase $\varphi_{d}$, has a mirror bound state with orientation
$-\beta$ that sees the ``--'' lobe of the $ d_{x^2-y^2} $-pair wave
function with phase $\varphi_{d}+\pi$.  As a result, in the total
current perpendicular to the interface, all odd harmonics cancel, and
the Josephson coupling is reduced.  The leading term is
$I_{\perp}\sim\sin(2\varphi)$~\cite{ZAGOS} and the stable ground state
with $I_{\perp}=0$ is at $\varphi=\pm\pi/2$ and, thus, breaks time
reversal symmetry.  The Josephson current parallel to the interface,
however, has contributions from the odd harmonics and to leading order
$I_{\parallel}\sim\sin(\varphi)$.  Remarkably, this parallel
contribution is {\em nonzero} in the ground state and constitutes a
spontaneous current.

Let us first consider this property of the SND junction on a
phenomenological level by means of GL theory.  We describe the
superconducting state by two OP's, $\eta_{s}$ (s-wave) and $\eta_{d}$
(d-wave), which correspond to the local pairing amplitudes.  The
corresponding GL free energy ${\cal F}$ has the general form,
\begin{eqnarray}\nonumber
        \frac{{\cal F}}{f_{0}}\!&=&\!\!\int\!\! d^3r {\Big [}
        \sum_{\mu=s,d} \{(\frac{T}
        {T_{c\mu}}-1)|\eta_{\mu}|^2+\beta_{\mu}|\eta_{\mu}|^{4}
        +\xi^{2}_{\mu}|{\bf \Pi}\eta_{\mu} |^2 \} \\\nonumber
        &&+\gamma_{1}|\eta_s|^{2}|\eta_{d}|^{2}+\frac{\gamma_2}{2}
        (\eta^{*2}_{s}\eta^{2}_{d}+\eta^{2}_{s}\eta^{*2}_{d})
        +\frac{(\nabla\times{\bf A})^{2}}{8\pi f_{0}}\\
        &&+\tilde{\xi}^{2}((\Pi_{x}\eta_{s})^{*}(\Pi_{x}\eta_{d})
        -(\Pi_{y}\eta_{s})^{*}(\Pi_{y}\eta_{d})+{\rm c.c.})]\;,
\end{eqnarray}
where $f_{0}$ is a free energy density, $T_{cs}$ and $T_{cd}$ are the
transition temperatures of $\eta_{s}$ and $\eta_{d}$, respectively,
and $\beta_{s,d}$, $\gamma_{1,2}$, $\xi_{s,d}$, and $\tilde{\xi}$ are
real coefficients ($\xi_{s,d}$ corresponds to the
zero-temperature coherence length).  These coefficients and the
transition temperatures are in general different in the three regions
of the SND-junction.  We use ${\bf \Pi}=\nabla-(2\pi i/\Phi_{0}){\bf
A}$, with vector potential ${\bf A}$ and flux quantum
$\Phi_{0}=hc/2e$.  To study the properties of the SND-junction we
minimize this free energy with respect to $\eta_{s,d}$ and ${\bf A}$.
Assuming homogeneity along the interface the problem reduces to one
spatial dimension which corresponds to the [1,1,0]-direction in the
coordinates used in ${\cal F}$ ($\hat{x}=\hat{a}$ and
$\hat{y}=\hat{b}$).  We call this direction $x'$ and the perpendicular
ones $y'$ and $z$.

\begin{figure}[hbt]
        \label{fig:op}
        \unitlength1cm
        \begin{picture}(6.3,6.5)
        \put(0,0){\psfig{figure=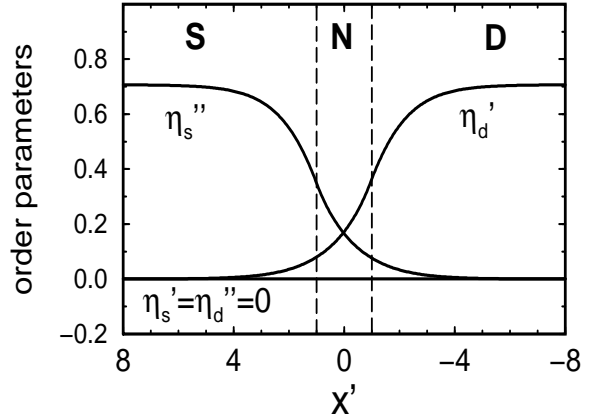,height=6.5cm,width=8.0cm}}
        \end{picture}
\caption{Spatial dependence of the OP in the SND-junction ($\alpha=
  \pi/4$) based on the GL theory.  The parameters of GL free energy
  are given in the text.  The temperature is $T=T_{cs,d}/2$ and the
  width $L$ of the normal metal layer is 2 in units of $\xi_{s}$.}
\end{figure}

We solve the complete set of GL equations numerically, for the case in
which the coefficients in ${\cal F}$ are identical for both OP's and
throughout the system.  The transition temperatures are only different
from zero in the corresponding superconducting regions.  We assume the
interfaces between the different layers to be completely transparent,
i.e. the OP's are continuous and have a continuous derivative.  For
our calculation we choose $\beta_{s}=\beta_{d}=1/2$,
$\xi_{s}=\xi_{d}=1$ (unit of length), $\gamma_{1}=4/5$,
$\gamma_{2}=2/5$ and $\tilde{\xi}=1$.  This leads to
$f_{0}=H^{2}_{c}/8\pi$, where $H_{c}$ is the thermodynamic critical
field at $T=0$.  We fix $\Phi_{0}/2\sqrt{2}\pi H_{c}\xi^2_{s}=4$ which
corresponds to the London penetration depth $ \lambda $ at $ T=0$ in
units of $\xi_{s}$.  The result is shown in Fig.~2 for the OP's and in
Fig.~3 for the magnetic field and the supercurrent along the
$y'$-direction.

Both OP components penetrate the normal metal layer (proximity effect)
and coexist there in a combination, which for the case $\alpha=\pi/4$
is entirely determined by the mixing terms
$(\gamma_{2}/2)(\eta^{*2}_{s}\eta^{2}_{d}+\eta^{2}_{s}\eta^{*2}_{d})$.
Within the weak coupling approach which we assume to apply, at least,
within the normal metal layer, $\gamma_{2}$ is
positive~\cite{XU}. This term yields the basic
$\cos(2\varphi)$-dependence of the SND-junction free energy.  It fixes
the phase difference between $\eta_{s}$ and $\eta_{d}$ to $\varphi=
\varphi_{d}-\varphi_{s}= \pm\pi/2$ in accordance with the argument
given above.  The mixed state has the $ {\cal T} $-violating $s\pm
id$-character in the normal metal.

The supercurrent density follows from $ {\cal F} $ as ${\bf J}=
-2c\partial{\cal F}/\partial {\bf A}$.  We find that the current
component $J_{x'}=J_{\perp}$ vanishes in the stable junction state and
that a spontaneous supercurrent flows parallel to the $y'$-direction
and generates a magnetic field distribution $B_{z}$ in and close to
the metal layer (Fig.~3).  Within the GL-formulation the supercurrent
$J_{y'}=J_{\parallel}$ is caused by the spatial variation of the two
OP components,
\begin{equation}
        J_{y'}=\frac{\pi c\tilde{\xi}^{2}}{\Phi_{0}}{\rm Im}\{\eta_{s}
        \partial_{x'}\eta^{*}_{d}+\eta_{d}\partial_{x'}\eta^{*}_{s}\}\;,
\end{equation}
where we have omitted the diamagnetic part. Note that this part of $
J_{y'} $ has essentially the $ \sin \varphi $-dependence anticipated
above. Under symmetric conditions, $J_{y'}$ depends only weakly on
$x'$ inside the normal metal layer as shown in Fig.~3.  The induced
magnetic field is screened perpendicular to the interface on the scale
of the London penetration depth in the superconducting regions by
currents flowing in the opposite direction.

\begin{figure}[hbt]
        \label{fig:jb}
        \unitlength1cm
        \begin{picture}(6.3,6.5)
        \put(0,0){\psfig{figure=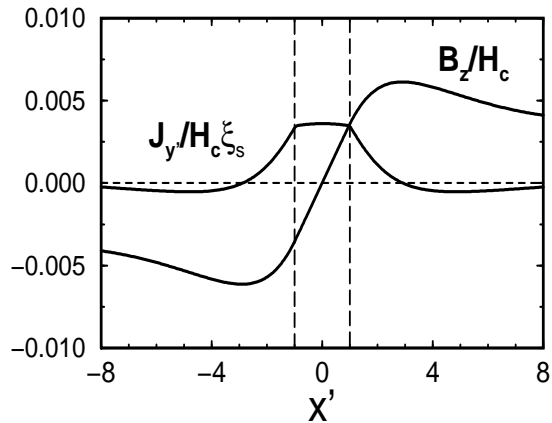,height=6.5cm,width=8.0cm}}
        \end{picture}
\caption{Spatial dependence of the spontaneous supercurrent and the
magnetic field in the SND junction based on the GL theory under the
same conditions as in Fig.~2}
\end{figure}

Let us turn now to the microscopic view by considering the bound state
solutions to the Bogoliubov-de Gennes equation in the normal metal
layer~\cite{KULIK} under the symmetric condition, i.e. the d-wave
energy gap in D has the form $\Delta_{d}= |\Delta|\rm{sign}
(\cos[2(\theta-\alpha)])$, with the amplitude $|\Delta|$ equal to that
of the gap of the s-wave superconductor in S.  We take the Fermi
momenta in S, N, and D to be equal and the transparency of the
interfaces to be high.  Furthermore, we also neglect the suppression
of the energy gap near the normal metal and assume the pairing
interaction to be zero in N.

\begin{figure}[hbt]
        \label{fig:ground}
        \unitlength1cm
        \begin{picture}(6.3,6.5)
        \put(-0.5,0){\psfig{figure=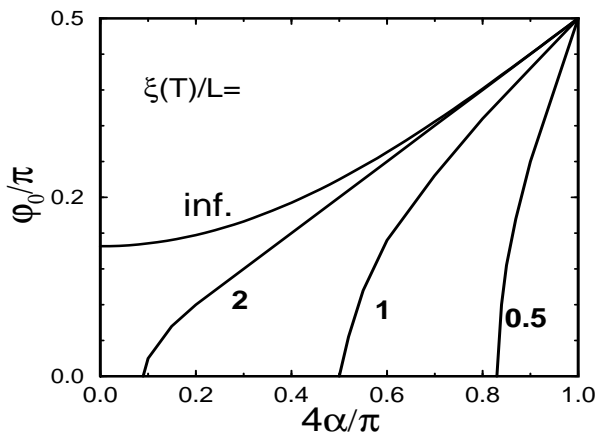,height=6.5cm,width=9cm}}
        \end{picture}
\caption{The ground state phase difference $\pm\varphi_{0}$ as a
  function of orientation angle $\alpha$ for temperatures
  corresponding to $\xi(T)/L=\infty$, 2, 1, and 0.5~.}
\end{figure}

The total Josephson current is a sum over all possible (bound) states
near the Fermi energy.  If the width of the normal metal $L$ is
smaller than the thermal coherence length $\xi_{T}$ and the elastic
mean free path $l$ in N, the Josephson current is given by~\cite{KZSJ},
\begin{equation}
        {\bf J}=\int dk_{y}dk_{z}\frac{2e{\bf k}_{F}}
        {m\pi L}\sum^{\infty}_{n=1}\frac{(-1)^{n+1}}{n}
        f_{n,\hat{\bf k}_{F}}\sin[n\varphi_{\hat{\bf k}_{F}}]\;,
\end{equation}
with the Fermi momentum ${\bf k}_{F}=(k_{x},k_{y},k_{z})$ and
$k^{2}_{x} +k_{y}^{2} +k_{z}^{2}\approx k^{2}_{F}$.  The integral runs
over all transverse momenta $|k_{y}|,|k_{z}|\leq k_{F}$ and the sum
over all possible numbers of multiple Andreev reflections $n$.  The
factors $f_{n,\hat{\bf k}_{F}}$ take the suppression due to thermal
decoherence and impurity scattering into account,
\begin{equation}
        f_{n,\hat{\bf k}_{F}}=\exp(-2nL_{\hat{\bf k}_{F}}/l)
        \frac{nL_{\hat{\bf k}_{F}}/
        \xi_{T}}{\sinh[nL_{\hat{\bf k}_{F}}/\xi_{T}]}\;,
\end{equation}
where we have introduced the normal metal coherence length $\xi_{T}=
\hbar v_{F}/(2\pi k_{B}T)$ in the clean limit, the mean free path $l$,
and the effective thickness of the normal metal layer $L_{\hat{\bf
k}_{F}} =Lk_{F}/(k^{2}_{F}-k^{2}_{y}-k^{2}_{z})^{1/2}$.

\begin{figure}[hbt]
        \label{fig:free}
        \unitlength1cm
        \begin{picture}(5.8,6.2)
        \put(-0.5,0){\psfig{figure=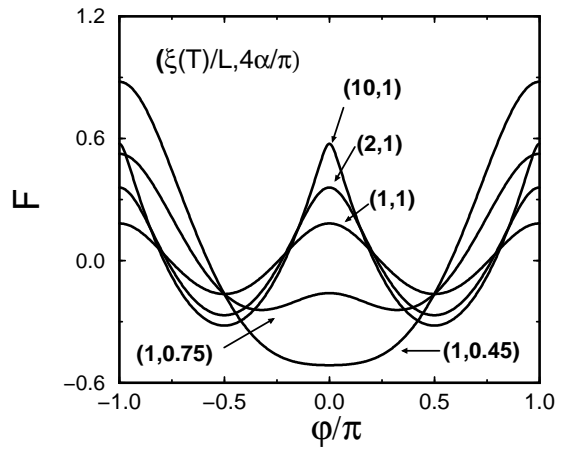,height=6.5cm,width=9cm}}
        \end{picture}
\caption{The junction free energy F as a function of the phase for
  $\alpha$=$\pi/4$ and $\xi_{T}/L$= 10, 2, 1, and for $\alpha$=0.45
  $\pi/4$ (one minimum) and 0.75$\pi/4$ (two shallow minima) at
  $\xi_{T}/L$=1.}
\end{figure}

The simplest case is that of zero temperature in the absence of
impurities, so that $\xi_{T}=l=\infty$ and all $f_{n}\equiv 1$.  In
this limit the sums over $n$ give sawtooth functions of the phase
difference, ${\rm saw}[\varphi_{\hat{\bf k}_{F}}]=[\varphi_{\hat{\bf
k}_{F}}+\pi] {\rm mod}2\pi$.  We obtain the Josephson currents
perpendicular and parallel to the junction immediately by angular
integration,
\begin{eqnarray}\nonumber
        I_{\perp}&=&A_{\perp} J_{0}[(\frac{\pi}{2}+\cos(2\alpha))
        {\rm saw}(\varphi)\\ &&\;\;\;\;\;\;\;\;\;\;\;\;\,
        +(\frac{\pi}{2}-\cos(2\alpha)){\rm saw}(\varphi+\pi)]\\
        I_{\parallel}&=&A_{\parallel} J_0\sin(2\alpha)[-{\rm saw}
        (\varphi)+{\rm saw}(\varphi+\pi)]\;.
\end{eqnarray}
Here $A_{\perp}$ and $A_{\parallel}$ denote the perpendicular and
parallel cross-section of the junction, and $J_{0}= ek^{3}_{F}/(\pi
mL)$.  Note that the current density $J_{0}$ is inversely proportional
to $L$, as in the GL calculation. 
The junction free energy $F(\varphi)$ is found by integrating $
I_{\perp} $ with respect to the phase.  It has two degenerate minima
at phase differences $\varphi_{0}=\pm[\pi/2 -\cos(2\alpha)]$, which
correspond to a parallel current along the junction $I_{\parallel} =
\pm BJ_0\sin(2\alpha)$.  The ground state has
$(s+e^{i\varphi_{0}}d)$-character in the normal metal layer as in the
phenomenological treatment, again reflecting ${\cal T}$-violation.

\begin{figure}[hbt]
        \label{fig:curr}
        \unitlength1cm
        \begin{picture}(6.3,6.7)
        \put(-1.5,0){\psfig{figure=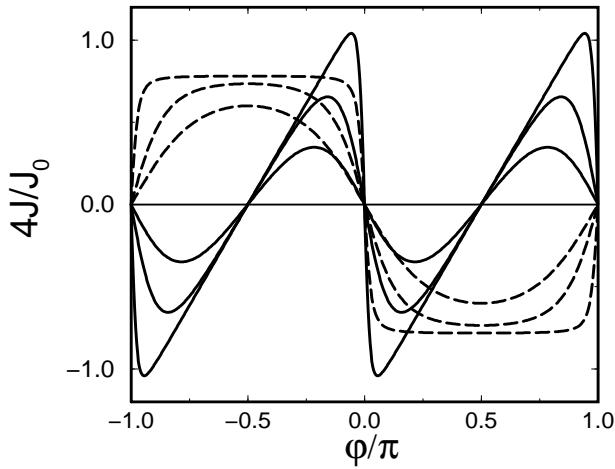,height=7.0cm,width=10cm}}
        \end{picture}
\caption{The parallel (dashed line) and perpendicular (solid line)
  Josephson current densities as a function of phase difference for
  $\alpha=\pi/4$ and temperatures corresponding to $\xi_{T}/L=$ 10,2,1
  (decreasing amplitude).}
\end{figure}

For nonzero temperature and in the presence of impurities, we evaluate
$I_{\perp}$, $I_{\parallel}$, and the junction free energy F
numerically.  In Fig.~4 the equilibrium phase difference $\varphi_{0}$
across the junction is plotted as a function of orientation angle
$\alpha$ for different temperatures $\xi_{T}/L$ in the case
$l=\infty$.  We find that time reversal symmetry is broken
($\varphi_{0} \neq 0, \pm\pi$) only for low enough temperatures, or
for the orientation angle $\alpha$ exceeding a critical value. For
$\alpha=\pi/4$, however, $\varphi_{0}= \pm\pi/2$ for all
$T<T_{cs},T_{cd}$ as in the GL treatment. The resulting phase diagram
is completely consistent with the one found by GL theories \cite{SK}.
The result for the junction free energy F is plotted in Fig.~5, and
$I_{\perp}$ and $I_{\parallel}$ in Fig.~6.  Both temperature and
disorder smear the sharp sawtooth structures found at $T=0$ in the
clean limit in a similar fashion.

The arbitrary equilibrium phase difference leads to experimentally
observable effects. $ {\cal T} $-violating junctions can lead to phase
windings which are non-integer multiples of $\pi$, giving rise to
non-standard (not (half-) integer) flux quantization. Thus, it is
possible to create devices including ${\cal T}$-violating junctions
which generate a spontaneous arbitrary magnetic flux
\cite{SBL,SK}. The observation of such a deviation from standard flux
quantization is a clear sign of ${\cal T}$-violation.  Furthermore,
the presence of two degenerate equilibrium states allows for
hysteresis effects ($ \varphi_{0} \leftrightarrow - \varphi_{0}$). By
applying a current through the junction one can switch between the two
states. This effect corresponds to a phase slip with a fractional flux
moving along the junction~\cite{SK}. This leads to dissipation and the
enhancement of microwave absorption as soon as the junction enters the
${\cal T}$-violating phase. The direct observation of the
spontaneous currents $I_{\parallel}$ or the field might be difficult,
since they average to zero over rather small length scales (London
penetration depth). Thus a probe with high spatial resolution would be
needed.

In summary, we have demonstrated that the Andreev bound states in the
normal metal layer of an SND-junction are the microscopic realization
of local ${\cal T}$-violation and provide a clear understanding of the
spontaneous current found in the phenomenological Ginzburg-Landau
analysis.  This observation allows for a more quantitative
consideration of this effect, which will be important for future
experimental investigations.  The experiments discussed at the end are
two among several possibilities to observe this ${\cal T}$-violating
state of the SND-junction.  Finally, we like to emphasize that this
effect is only possible in connection with unconventional
superconductivity and cannot occur for standard SNS-junctions.
Therefore, high temperature superconductivity provides an exciting new
class of Josephson phenomena.

We thank G. Blatter, V. Geshkenbein, and T.M. Rice for
helpful discussions.  This work was supported by the Swiss
Nationalfonds (PROFIL fellowship, M.S.) and the National Science
Foundation (DMR 95-28535, A.v.O.). A.H. thanks R. Scalettar and
G. Zimanyi at the UC Davis physics department for their hospitality.

\vspace{1cm}

$*$ To whom correspondence should be addressed.

$\dagger$ E-mail: avo@solid.ucdavis.edu.

\end{document}